\documentclass[a4paper,dvips,12pt]{article}
\usepackage{cite}
\usepackage{epsf}
\usepackage{multirow}
\usepackage{bm,bbm} 
\usepackage[pdftex]{graphicx}
\usepackage{tabularx}
\usepackage{latexsym}
\usepackage{amsmath,amssymb,exscale}
\usepackage{array,multicol}

\def\draftlabel#1{{\@bsphack\if@filesw {\let\thepage\relax
   \xdef\@gtempa{\write\@auxout{\string
      \newlabel{#1}{{\@currentlabel}{\thepage}}}}}\@gtempa
   \if@nobreak \ifvmode\nobreak\fi\fi\fi\@esphack}
        \gdef\@eqnlabel{#1}}
\def\@eqnlabel{}
\def\@vacuum{}
\def\draftmarginnote#1{\marginpar{\raggedright\scriptsize\tt#1}}
\def\draft{\oddsidemargin -.5truein
        \def\@oddfoot{\sl preliminary draft \hfil
        \rm\thepage\hfil\sl\today\quad\militarytime}
        \let\@evenfoot\@oddfoot \overfullrule 3pt
        \let\label=\draftlabel
        \let\marginnote=\draftmarginnote
   \def\@eqnnum{(\theequation)\rlap{\kern\marginparsep\tt\@eqnlabel}%
\global\let\@eqnlabel\@vacuum}  }

\newcommand{\PRL}[3]{\emph{ Phys.~Rev.~Lett.} \textbf{#1} (#2) #3}

\newcommand{\PR}[3]{\emph{ Phys.~Rep.} \textbf{#1} (#2) #3}

\def\ov{\overline}

\def\dalemb#1#2{{\vbox{\hrule height .#2pt
         \hbox{\vrule width.#2pt height#1pt \kern#1pt
                 \vrule width.#2pt}
         \hrule height.#2pt}}}

\def\half{{\textstyle{1\over2}}}
\let\a=\alpha    
    
    \let\p=\pi 
\let\s=\sigma     

      \let\G=\Gamma  
  \let\S=\Sigma   
\let\F=\Phi

 \def\bd{\begin{document}} \def\ed{\end{document}}
\def\ds{\documentstyle} \let\fr=\frac \let\bl=\bigl \let\br=\bigr
\let\Br=\Bigr \let\Bl=\Bigl
\let\bm=\bibitem
\let\na=\nabla
\let\pa=\partial
\let\ov=\overline
\def\ie{{\it i.e.\ }}
\def\tr{{\mbox{\rm tr}}}
\def\simlt{\mathrel{\lower2.5pt\vbox{\lineskip=0pt\baselineskip=0pt
            \hbox{$<$}\hbox{$\sim$}}}}
\def\simgt{\mathrel{\lower2.5pt\vbox{\lineskip=0pt\baselineskip=0pt
            \hbox{$>$}\hbox{$\sim$}}}}
\def\A{{\cal A}}
\def\a{{\mathcal a}}
\def\V{{\cal V}}
\def\F{{\cal F}}
\def\p{{\mathcal \phi}}
\def\L{{\mathcal L}}
\def\M{{\mathcal M}}
\def\bD{{\ov {\rm D}}}
\def\bO{{\ov {\rm O}}}
\def\bOp{{\ov {\rm O'}}}
\def\O{{ {\rm O}}}

\def\Om{\Omega}
\def\om{\omega}
\def\th{\theta}
\def\vt{\vartheta}
\def\vphi{\varphi}
\def\inte{{\bf Z}}
\def\pim{{\rm Im\,}}
\def\cn{{\cal N}}
\def\cpx{{\bf C}}
\def\real{{\bf R}}

\newcommand{\nsect}{\setcounter{equation}{0}
\def\theequation{\thesection.\arabic{equation}}\section}
\newcommand{\nappend}{\setcounter{equation}{0}
\def\theequation{\rm{A}.\arabic{equation}}\section*}
\newcommand{\appendixA}{\setcounter{equation}{0}
\def\theequation{\rm{A}.\arabic{equation}}\section*}
\newcommand{\appendixB}{\setcounter{equation}{0}
\def\theequation{\rm{B}.\arabic{equation}}\section*}
\newcommand{\appendixC}{\setcounter{equation}{0}
\def\theequation{\rm{C}.\arabic{equation}}\section*}
\newcommand{\appendixD}{\setcounter{equation}{0}
\def\theequation{\rm{D}.\arabic{equation}}\section*}
\newcommand{\appendixE}{\setcounter{equation}{0}
\def\theequation{\rm{E}.\arabic{equation}}\section*}
\newcommand{\appendixF}{\setcounter{equation}{0}
\def\theequation{\rm{F}.\arabic{equation}}\section*}
\newcommand{\appendixG}{\setcounter{equation}{0}
\def\theequation{\rm{G}.\arabic{equation}}\section*}
\def\baselinestretch{1.5}
\catcode`\@=11
\def\marginnote#1{}
\newcount\hour
\newcount\minute
\newtoks\amorpm
\hour=\time\divide\hour by60
\minute=\time{\multiply\hour by60 \global\advance\minute by-\hour}
\edef\standardtime{{\ifnum\hour<12 \global\amorpm={am}%
        \else\global\amorpm={pm}\advance\hour by-12 \fi
        \ifnum\hour=0 \hour=12 \fi
        \number\hour:\ifnum\minute<10 0\fi\number\minute\the\amorpm}}
\edef\militarytime{\number\hour:\ifnum\minute<10 0\fi\number\minute}
\def\draftlabel#1{{\@bsphack\if@filesw {\let\thepage\relax
   \xdef\@gtempa{\write\@auxout{\string
      \newlabel{#1}{{\@currentlabel}{\thepage}}}}}\@gtempa
   \if@nobreak \ifvmode\nobreak\fi\fi\fi\@esphack}
        \gdef\@eqnlabel{#1}}
\def\@eqnlabel{}
\def\@vacuum{}
\def\draftmarginnote#1{\marginpar{\raggedright\scriptsize\tt#1}}
\def\draft{\oddsidemargin -.5truein
        \def\@oddfoot{\sl preliminary draft \hfil
        \rm\thepage\hfil\sl\today\quad\militarytime}
        \let\@evenfoot\@oddfoot \overfullrule 3pt
        \let\label=\draftlabel
        \let\marginnote=\draftmarginnote
   \def\@eqnnum{(\theequation)\rlap{\kern\marginparsep\tt\@eqnlabel}%
\global\let\@eqnlabel\@vacuum}  }

\def\preprint{\twocolumn\sloppy\flushbottom\parindent 1em
        \leftmargini 2em\leftmarginv .5em\leftmarginvi .5em
        \oddsidemargin -.5in    \evensidemargin -.5in
        \columnsep 15mm \footheight 0pt
        \textwidth 250mmin      \topmargin  -.4in
        \headheight 12pt \topskip .4in
        \textheight 175mm
        \footskip 0pt
\def\@oddhead{\thepage\hfil\addtocounter{page}{1}\thepage}
        \let\@evenhead\@oddhead \def\@oddfoot{} \def\@evenfoot{} }

\def\titlepage{\@restonecolfalse\if@twocolumn\@restonecoltrue\onecolumn
     \else \newpage \fi \thispagestyle{empty}\c@page\z@ 
        \def\thefootnote{\fnsymbol{footnote}} }

\def\endtitlepage{\if@restonecol\twocolumn \else  \fi
        \def\thefootnote{\arabic{footnote}}
        \setcounter{footnote}{0}}  

\catcode`@=12
\relax
\def\abs#1{\left| #1\right|}
\def\bC{\mathop{\bf C}}
\def\bea{\begin{array}}
\def\bem{\begin{displaymath}}
\def\beq{\begin{equation}}
\def\beqa{\begin{eqnarray}}
\def\bR{\mathop{\bf R}}
\def\bra#1{\left\langle #1\right|}
\def\eea{\end{array}}
\def\eem{\end{displaymath}}
\def\eeq{\end{equation}}
\def\eeqa{\end{eqnarray}}
\def\eq{\beq\eeq}                          
\def\eqr#1{\beq\label#1\eeq}               
\def\half{\frac{1}{2}}
\def\Im{\mathop{\rm Im}}
\def\ket#1{\left| #1\right\rangle}
\def\sket#1{| #1 >}
\def\lie{\hbox{\it \$}}                          
\def\lineint{\oint \frac{d z}{2 \pi i}} 
\def\modsq#1{| #1 |^2}
\def\NP#1#2#3{Nucl. Phys. \underline{#1} (19#2) #3}
\def\ov{\overline}
\def\partder#1#2{{\partial #1\over\partial #2}}
\def\PL#1#2#3{Phys. Lett. \underline{#1} (19#2) #3}
\def\PR#1#2#3{Phys. Rev. \underline{#1} (19#2) #3}
\def\PRL#1#2#3{Phys. Rev. Lett. \underline{#1} (19#2) #3}
\def\Re{\mathop{\rm Re}}
\def\secder#1#2#3{{\partial^2 #1\over\partial #2 \partial #3}}
\def\s2w{\sin^2 \theta_W}
\def\Tr{\mathop{\rm Tr}}
\def\und{\underline}
\def\VEV#1{\left\langle #1\right\rangle} \let\vev\VEV
\def\mbf#1{\hbox{\boldmath $#1$}}
\relax
\def\dalpha{{\dot\alpha}}
\def\dbeta{{\dot\beta}}
\def\drho{{\dot\rho}}
\def\dsigma{{\dot\sigma}}
\def\crbig{\\\noalign{\vspace {3mm}}}
\def\bigint{{\displaystyle\int}}
\def\S{\Sigma}
\def\G{\Gamma}
\def\L{{\cal L}}
\def\SG{S_{\Gamma}}
\def\Fint{{\bigint d^2\theta\,}}
\def\Fbarint{{\bigint d^2\ov\theta\,}}
\def\Dint{{\bigint d^2\theta d^2\ov\theta\,}}

\title{ \vspace*{-1.8cm}
\vspace{1cm}
\bf{$w_{\infty}$ $3$-algebra} \vspace*{-0.3cm}}

\begin{document}
\author{\bf\large{
Shankhadeep Chakrabortty
\footnote{sankha@iopb.res.in}~,
Alok Kumar\footnote{kumar@iopb.res.in}~, 
Sachin Jain\footnote{sachjain@iopb.res.in}}\\  
\\[-3mm]
\emph{\normalsize Institute of Physics, Bhubaneswar 751 005, 
India}\\
}

\date{\today}

\maketitle
\thispagestyle{empty}

\begin{abstract}
We use `lone-star' product of the $W_{\infty}$ generators as well as their 
commutation relations to obtain a $w_{\infty}$ 3-algebra by applying 
appropriate double scaling limits on the generators.
We show explicitly that "Fundamental Identity condition" (FI) of 3-algebra is satisfied.
\end{abstract}
\date

\vspace*{-0.8cm}
\newpage

3-Algebra relations have played an important role in the 
construction of the worldvolume theories of multiple $M2$ branes
\cite{BL,Gustav,matsuo}, which has attracted a great deal of 
attention. Of particular 
interest are the developments in the realization of this structure 
in terms of three dimensional $N=6$ Chern-Simons 
theories\cite{malda}. These pioneering investigations 
have also generated renewed interest in the analysis of  
3-algebras\cite{nambu,zachos0} in general.
In particular, several authors\cite{Lin,Zachos,Lars,Soch} 
have considered their affine extensions
by showing the existence of Kac-Moody and (centerless) Virasoro 
3-algebras and demonstrating  some of their applications to the 
Bagger-Lambert theory.  

Motivated by the above recent works, specially those in 
\cite{Lin,Zachos,Lars,Soch}, as well as by the fact that 3-algebra relations
are a rarity \cite{matsuo},  
in this paper, we explicitly obtain a
classical $w_{\infty}$ 3-algebra and show that our relation satisfies the 
3-algebra "Fundamental Identity condition"(FI). Our construction is based on 
earlier work on $W_{\infty}$ and $W_{1+\infty}$ symmetries
(see \cite{pope,pope1} {\it and references therein}). Using the 
`lone-star' product of $W_{1+\infty}$ generators and their commutation 
relations we write down a 3-algebra relation. The structure constants for
such a 3-algebra relation simplify in a double scaling limit and we show 
the validity of the FI through direct verification presented below.
We also show that the 4-brackets among the resulting $w_{\infty}$
generators vanish.

We now start with the commutation relations defining $W_{1+\infty}$
algebra written in terms of generators $\tilde{V}^i_m$ \cite{pope}:
\beq
   [\tilde{V}^i_m, \tilde{V}^j_n] = \sum_{l\geq 0} q^{2l} 
		\tilde{g}^{ij}_{2l}(m, n)\tilde{V}^{i+j-2l}_{m+n}
	+ q^{2i}\tilde{c}_i(m)\delta^{ij}\delta_{m+n, 0} , 
\label{W-infty-algebra}
\eeq
where  superscripts $i,j,l$, representing the conformal spin of the 
generators, are in general integers: $-1, 0, 1, \cdots$ etc. 
whereas integer subscripts $m, n$ can take arbitrary positive or negative
values. We also have:
\beq
   	\tilde{g}^{ij}_{l}(m, n) \equiv 
		{g}^{ij}_{l}(m, n, -\frac{1}{2}) 
\label{tilde-g}
\eeq
given by an expression:
\beq
{g}^{ij}_{l}(m, n, s) = \frac{1}{2(l+1)!}{\phi}^{ij}_{l}(s)
				{N}^{ij}_{l}(m,n) .
\eeq
Explicitly, $\phi^{ij}_l (s)$ are given by a generalized hypergeometric
function:
\beq
	{\phi}^{ij}_{l}(s) = {}_4F_3\left[  \begin{array}{ccccc}
-\frac{1}{2}-2s , &  \frac{3}{2}+ 2s,& - \frac{l}{2}-\frac{1}{2},
& - \frac{l}{2}& \\
&  & & & ; 1\\
-i -\frac{1}{2}, &  - j -\frac{1}{2}, & i+j-l+\frac{5}{2}& &
			\end{array}\right]
\label{phi-ij}	
\eeq
and 
\beq
 N^{ij}_l (m, n) = \sum_{k=0}^{l+1} (-1)^k 
\begin{pmatrix} l+1 \\ k \end{pmatrix}
(2i + 2 - l)_k [2j+2-k]_{l+1-k} [i+1+m]_{l+1-k}[j+1+n]_k ,
\label{N} 
\eeq
where $[a]_n$ stands for $\frac{a!}{(a-n)!}$ and 
$(a)_n$ stands for $\frac{(a+n-1)!}{(a-1)!}$. Also,
in eq. (\ref{W-infty-algebra}) $q$ is an arbitrary scaling parameter,
which we will fix later on through a double scaling process. Finally,
the central term of the algebra in eq.  (\ref{W-infty-algebra})
can be consistently set to zero and corresponds to the analysis 
of classical symmetries.

Another property of interest for us will be the `lone-star' product
of the $W_{1+\infty}$ generators:
\beq
\tilde{V}^i_m \star \tilde{V}^j_n = \sum_{l\geq -1} q^{l} 
		\tilde{g}^{ij}_{l}(m, n)\tilde{V}^{i+j-l}_{m+n} .
\label{lone-star}
\eeq
This star product is classical, since it does not contain information
about the central term. As in the following we make use of the relation
(\ref{lone-star}) to construct our 3-algebra, this analysis therefore holds
for the classical case only. Note also that the commutation relation
(\ref{W-infty-algebra}) follows from the `lone-star' product 
eq. (\ref{lone-star}) (in absence of central term)
by realizing that coefficients 
$\tilde{g}^{ij}_l (m, n)$ are symmetric under the simultaneous interchange of
$i, j$ and $m, n$ for odd $l$'s whereas they are antisymmetric for 
even $l$'s. We now restrict ourselves to the case when the central term is
absent.

Now, using the definition of the 3-algebra relation:
\beq
    [A, B, C] = A [ B, C] + B [C, A] + C[A, B]
\label{def-3-algebra}
\eeq
and the commutation relation (\ref{W-infty-algebra}) as well as the
star product (\ref{lone-star}), we can write the 3-algebra relation:
\beqa
   [\tilde{V}^a_m, \tilde{V}^b_n, \tilde{V}^c_p] 
= \sum_{l\geq 0, r\geq -1} q^{2l+r} [
 \tilde{g}^{b, c}_{2l}(n, p)\tilde{g}^{a, b+c-2l}_{r}(m, n+p) + \cr
\tilde{g}^{c, a}_{2l}(p, m)\tilde{g}^{b, c+a-2l}_{r}(n, m+p) +
\tilde{g}^{a, b}_{2l}(m, n)\tilde{g}^{c, a+b-2l}_{r}(p, m+n)]
\tilde{V}^{a+b+c-2l-r}_{m+n+p} ,
\label{W-3-algebra1}
\eeqa
where the index $r$, for a given $l$, runs over indices 
$r = -1, 0, \cdots, (a+b+c-2l+1)$, whereas the running of index
$l$ in the three terms in rhs of eq. (\ref{W-3-algebra1}) is from  
zero upto $b+c$, $c+a$ and $a+b$ respectively.

The 3-algebra relation in eq. (\ref{W-3-algebra1})
may be of interest in its own right,
however in the following we present a  simpler situation 
by using a double scaling limit on the above relation. We also recall
that a similar procedure (but with a single scaling parameter $q$)
was used earlier to obtain the $w_{\infty}$-algebra from $W_{\infty}$.
Relationship between $w_{\infty}$-algebra and area preserving 
reparameterizations of 2-torus are also well known \cite{zachos2}. 
In this paper we will observe an interesting relation
at the 3-algebra level by comparing the structure constants of the 3-algebra
emerging from the 3-bracket given in eq. (\ref{W-3-algebra1}), after 
taking the double scaling limit, with 
the one for the classical Nambu 3-brackets of 
globally defined functions\footnote{This point was communicated to us by
Cosmos Zachos and collaborators.}
$f, g, h$ on $T^2$.

Now, to apply our double scaling, we scale all the generators $\tilde{V}^a_m$
in eq. (\ref{W-3-algebra1}) by a parameter $\beta$. Note that such a 
scaling is in addition to the one given in \cite{pope}
which lead to the powers of $q^{2l}$ in the commutation relation
(\ref{W-infty-algebra}). 
We also note that the smallest power of $q$
in eq. (\ref{W-3-algebra1}) corresponds to $l=0$ and $r=-1$.
In order to keep only this term, after the double scaling, 
we take the limits: $q \rightarrow 0$, $\beta \rightarrow \infty$
such that $\beta^2 q = 1$. We then obtain the simplified 3-algebra
in terms of the rescaled generators $w^a_m$'s:
\beq
   [w^a_m, w^b_n, w^c_p] 
= [c(n-m) + b (m-p) + a (p-n)] w^{a+b+c+1}_{m+n+p} ,
\label{w-3-algebra1}
\eeq
where we have also made use of the fact that
\beq
	\tilde{g}^{ab}_{-1}(m, n) = 1 ,\,\,\,\,
	\tilde{g}^{ab}_{0}(m, n) =  (b+1)m - (a+1)n .
\label{g0g-1} 
\eeq

We now verify that $w_{\infty}$ 3-algebra satisfies the FI, written in
the present case as:
\beqa
\hspace{-0.2in}
[w^a_m, w^b_n, [w^c_p, w^d_q, w^e_r] ] = 
[[w^a_m, w^b_n, w^c_p], w^d_q, w^e_r] ] + 
[w^c_p, [w^a_m, w^b_n, w^d_q],  w^e_r]] +
[w^c_p, w^d_q, [w^a_m, w^b_n,  w^e_r]] . 
\label{FI}
\eeqa

A discussion on the necessity of the FI's defining the Leibniz rule for 
the action of 3-brackets, as well as an analysis of the associativity 
constraints in such cases, is presented in \cite{Zachos}. In our case,
evaluating the four terms we obtain:
\beqa
[w^a_m, w^b_n, [w^c_p, w^d_q, w^e_r] ] = 
[e(q-p) + d (p-r) + c(r-q)] \times \hspace{2.0in}\cr
[(c+d+e+1)(n-m) + b (m-p-q-r) +
a(p+q+r-n)]w^{a+b+c+d+e+2}_{m+n+p+q+r} , 
\label{term1}
\eeqa
\beqa
[[w^a_m, w^b_n, w^c_p], w^d_q, w^e_r] ] = 
[c(n-m) + b (m-p) + a(p-n)] \times \hspace{2.0in}\cr
[e(q-m-n-p) + d (m+n+p-r) +
(a+b+c+1)(r-q)]w^{a+b+c+d+e+2}_{m+n+p+q+r} , 
\label{term2}
\eeqa
\beqa
[w^c_p, [w^a_m, w^b_n, w^d_q],  w^e_r]] = 
[d(n-m) + b (m-q) + a(q-n)] \times \hspace{2.0in}\cr
[e(m+n+q-p) + (a+b+d+1)(p-r) +
c(r-m-n-q)]w^{a+b+c+d+e+2}_{m+n+p+q+r} , 
\label{term3}
\eeqa
\beqa
[w^c_p, w^d_q, [w^a_m, w^b_n,  w^e_r]] = 
[e(n-m) + b (m-r) + a(r-n)] \times \hspace{2.0in}\cr
[(a+b+e+1)(q-p) + d(p-m-n-r) +
c(m+n+r-q)]w^{a+b+c+d+e+2}_{m+n+p+q+r} . 
\label{term4}
\eeqa
Using eqs. (\ref{term1}), (\ref{term2}), (\ref{term3}) and (\ref{term4}),
it can now be checked directly that the 3-algebra in
eq. (\ref{w-3-algebra1}) satisfies the FI in eq. (\ref{FI}).

We have therefore obtained a 3-algebra generalization of the 
$w_{\infty}$-algebra. Note that our double scaling is such that 
it gives a nontrivial 3-algebra in terms of $w_{\infty}$
generators. This double scaling would however make the 
original commutation relations\cite{pope} of $w_{\infty}$
generators trivial. There is, however, no inconsistency with
our analysis above, since the `lone-star' product also goes to 
infinity in this limit, thus giving us a  
well defined 3-algebra with finite coefficients.
We have also analyzed the  expressions for the 
(totally antisymmetrized) 4-brackets involving 
the generators $w^a_m$, using the relation (\ref{w-3-algebra1}):
\beq
[w^a_m, w^b_n, w^c_p, w^d_q] 
= w^a_m [w^b_n, w^c_p, w^d_q] - w^b_n[w^c_p, w^d_q, w^a_m] + 
w^c_p [w^d_q, w^a_m, w^b_n]- w^d_q [w^a_m, w^b_n, w^c_p] .
\label{4-bracket}
\eeq
By explicit calculation we find that it is identically zero, a result
similar to the one \cite{Zachos} for the Virasoro 3-algebra.

As already pointed out before, above results can also be reinterpreted in 
terms of the algebraic structure of the reparameterizations of 2-torus
through the evaluation of the classical Nambu 3-brackets (3CNB)
of globally defined functions $f, g, h$ on a 2-torus.
3CNB of functions $f,g,h$, that are completely antisymmetrized, are defined as
the Jacobian of the transformation from ($x, y, z$) to 
($f(x,y,z), g(x,y,z), h(x,y,z)$):
\beqa
\{ f, g, h \} \equiv \frac{\partial{(f, g, h)}}{\partial{(x, y, z)}} \equiv
\,\, f \{ g,h \} + g \{ h,f \}  + h \{ f,g \} \cr
= \frac { \partial  f }{ \partial  x} 
({ \frac { \partial  g }{ \partial  y} \frac { \partial  h }{ \partial  z } 
- \frac { \partial  h }{ \partial  y} \frac { \partial  g }{ \partial  z }}) 
+ \frac { \partial g }{ \partial  x} ({ \frac { \partial  h }{ \partial  y} 
\frac { \partial  f }{ \partial  z } - \frac { \partial  f }{ \partial  y} 
\frac { \partial  h }{ \partial  z }})  + 
\frac { \partial  h }{ \partial  x} ({ \frac { \partial f }{ \partial  y} 
\frac { \partial  g }{ \partial  z } - 
\frac { \partial  g }{ \partial  y} \frac { \partial f }{ \partial  z }}).
\label{pb}
\eeqa
Now, to establish the connection with our results given above, we note that
by choosing: 
\beqa
f \equiv w^{a}_{m} = \sqrt z \exp{((a + {\frac 12})x + my)},\cr
g \equiv w^{b}_{n} = \sqrt z \exp{((b + {\frac 12})x + ny)},\cr
h \equiv w^{c}_{p} = \sqrt z \exp{((c + {\frac 12})x + py)},
\label{nambu-fgh}
\eeqa
we obtain the 3CNB of generators $\{w^{a}_{m}, w^{b}_{n}, w^{c}_{p} \}$,
which matches with the 3-bracket given in eq. (\ref{w-3-algebra1}) (by 
a constant scaling of the generators),
with structure constant:
\beq
= {\frac 12}\left| \begin{array}{ccc}
1 & 1 & 1 \\
a & b & c \\
m & n & p \end{array} \right|.
\label{structure-constant}
\eeq
We also note that a somewhat similar structure appeared in  
the Moyal (sine) brackets of \cite{zachos2} and its 
correspondence to 3-algebra structure constants in our case
will be of interest to examine. Also, it is noticed from 
eq. (\ref{nambu-fgh}) that the 3-algebra generators of eq. 
(\ref{w-3-algebra1}) can be identified with the modes
of the deformations of 2-torus \cite{zachos2}. In the present case,
however, one also needs to multiply the exponential functions 
in eqs. (\ref{nambu-fgh}) by an extra factor $\sqrt{z}$ common 
to all three generators in the 3CNB. The geometric interpretation 
of such an extra factor may be possible by identifying the 
complete geometry as a direct product of 2-torus with a point,
since the deformation mode along the $z$ direction is frozen.

We now comment on the the validity of 
the 3-bracket expression (before taking the scaling limit),
i.e. eq. (\ref{W-3-algebra1}),
as a proper 3-algebra relation. First of all it is interesting to 
note that all the terms which are even in $r$, in the sum in the 
rhs of eq. (\ref{W-3-algebra1}), vanish. This follows from an 
observation of the Jacobi identity involving the $W_{1+\infty}$
generators $\tilde{V}^a_m$. Moreover, we have explicitly 
analyzed the FI for some of the low lying (i.e. in indices $a, b$)
generators $\tilde{V}^a_m$ and our results imply that $w_{\infty}$
does not extend to the full $W_{1+\infty}$. In other words, not all
3-brackets satisfy the FI, a situation already known for the case
of compact Lie group generators.  
Eq. (\ref{W-3-algebra1}) may, however, still be of interest in obtaining
a consistent higher bracket.

It is also of interest  
to generalize this result in several other directions, 
such as in obtaining `primary' field representations,
supersymmetric  generalizations etc.. These topics are currently 
under investigation.

\noindent {\bf{\large Acknowledgement}}

After the first submission of our paper to the archive, we had several 
useful communications on our 3-algebra relation with Cosmos Zachos and 
collaborators.  We would like to thank them for pointing out to us 
interesting connections (as already included in the text)
such as the ones in eqs. (\ref{pb}), (\ref{nambu-fgh})
between our 3-algebra eq. (\ref{w-3-algebra1}) and 
classical Nambu brackets as well as  $T^3$ diffeomorphisms.

\end {document}